\documentclass[conference]{IEEEtran}
\IEEEoverridecommandlockouts
\usepackage{cite}
\usepackage{enumitem}
\usepackage{amsmath,amssymb,amsfonts}
\usepackage{algorithmic}
\usepackage{graphicx}
\usepackage{multirow}
\usepackage{textcomp}
\usepackage{xcolor, colortbl}
\definecolor{LightCyan}{rgb}{0.88,1,1}
\def\BibTeX{{\rm B\kern-.05em{\sc i\kern-.025em b}\kern-.08em
    T\kern-.1667em\lower.7ex\hbox{E}\kern-.125emX}}
\begin{document}

\title{Survey of Malware Analysis through Control Flow Graph using Machine Learning \\
}

\author{\IEEEauthorblockN{Shaswata Mitra}
\IEEEauthorblockA{\textit{Department of Computer Science and Engineering} \\
\textit{Mississippi State University}\\
MS, USA \\
sm3843@msstate.edu}
\and
\IEEEauthorblockN{Stephen A. Torri}
\IEEEauthorblockA{\textit{Department of Computer Science and Engineering} \\
\textit{Mississippi State University}\\
MS, USA \\
storri@cse.msstate.edu}
\and
\IEEEauthorblockN{Sudip Mittal}
\IEEEauthorblockA{\textit{Department of Computer Science and Engineering} \\
\textit{Mississippi State University}\\
MS, USA \\
mittal@cse.msstate.edu}
}

\author{\IEEEauthorblockN{Shaswata Mitra\IEEEauthorrefmark{1}, Stephen A. Torri\IEEEauthorrefmark{2}, 
Sudip Mittal\IEEEauthorrefmark{3}
}
%\IEEEauthorblockA{
Department of Computer Science \& Engineering\\ Mississippi State University
\\ sm3843@msstate.edu\IEEEauthorrefmark{1}, storri@cse.msstate.edu\IEEEauthorrefmark{2},
mittal@cse.msstate.edu\IEEEauthorrefmark{3}\\
}%}

\maketitle

\begin{abstract}
    
    Malware is a significant threat to the security of computer systems and networks that requires sophisticated techniques to analyze its behavior and functionality for detection. Due to their rapid evolution, traditional signature-based malware detection methods have become ineffective in detecting new and unknown malware. One of the most promising techniques that can overcome the limitations of signature-based detection is to use control flow graphs (CFGs). CFGs leverage the structural information of a program to represent the possible paths of execution as a graph, where nodes represent instructions and edges represent control flow dependencies. Machine learning (ML) algorithms extract these features from CFGs and classify them as malicious or benign. In this survey, we aim to review some state-of-the-art methods for malware detection through CFGs using ML, focusing on the different ways of extracting, representing, and classifying. Specifically, we present a comprehensive overview of different types of CFG features used and different ML algorithms applied to CFG-based malware detection. We provide an in-depth analysis of the challenges and limitations of these approaches, as well as suggest potential solutions to address persisting open problems and promising future directions for research in this field. 

\end{abstract}


\section{Introduction}

    %  --- Introduction Points --- 
    % What was the research topic being investigated?
    % Why was this topic important to study?
    % What did we know about this topic before I did this study? The background knowledge the reader requires to understand the rest of the paper.
    % How will this study advance the understanding of computer science?
    
    Malware is malicious software designed to damage or gain unauthorized access to computer systems. It is a significant threat to computer systems, causing billions of dollars in yearly damages. To maintain a secured cyber-space— malware detection and analysis, therefore, have become extremely important with the increased presence of malware and cyber-attacks every day. To detect and analyze malware, researchers use various static and dynamic techniques. Traditionally, malware detection is done using a static approach, where program hash signatures are compared to identify malware presence. Due to the recent development of numerous signature spoofing techniques, the hash-comparison technique has seen reduced effectiveness. One alternative technique used to analyze malware is through a control flow graph (CFG). CFG analysis is a powerful approach used in computer science to determine the behavior of programs. It is a graphical representation of the execution flow of a program, which can be used to identify abnormal patterns and malicious behavior in the program. In cybersecurity, CFG analysis has become a critical static analysis technique for malware detection. 
    
    The importance of CFG analysis in malware detection lies in its ability to provide a meticulous view of program execution, allowing security analysts to understand the program's logic, identify potential vulnerabilities, and detect the presence of malicious behaviors, as it can reveal hidden or obfuscated code, and expose malicious behavior that would otherwise go undetected with other static approaches. This technique has been widely used in cybersecurity, and its effectiveness has been demonstrated in numerous studies. Due to the requirement of thorough analysis by professional security analysts and limited automation scopes, such an approach used to be cost- and time-prohibitive. However, recent advancements in machine learning (ML), deep learning (DL), and data analysis have enabled sophisticated and accurate analysis of CFGs for malware detection in an automated, timely, and cost-effective way. 
    
    In this study, we explore in detail the recent advancements of CFG analysis through ML in detail, its use cases in malware detection, persisting drawbacks, and further improvement areas. To fully grasp the topic, readers will require a basic understanding of programming concepts, cybersecurity fundamentals, and ML. By providing a comprehensive overview of CFG analysis in malware detection, this study will contribute to the ongoing efforts using ML to enhance cybersecurity and protect against emerging threats.
    
    In section \ref{research_method}, we discuss the research objective and criteria that shape the scope of the survey. In section \ref{results}, we discuss the primary research findings that adhere to the rules set in section \ref{research_method}. Finally, in section \ref{discussion}, we address the research questions set in \ref{research_questions} with drawbacks and future recommendations.

\section{Research Method}
\label{research_method}

    Our study aims to address recent developments in cybersecurity to analyze and identify malware through CFG analysis using ML. It is not a full-fledged systematic literature review (SLR) set forth by Kitchenham et al. \cite{kitchenham2004procedures} covering all the developmental works. Instead, we aim to address some popular different ML frameworks that shaped the present research landscape. Furthermore, we also focus on providing a preliminary understanding of the topic with relevant literature. 
    
    Hence, we followed the procedures outlined in the following sub-sections to conduct our review. This approach allowed us to provide an overview of the current research landscape while acknowledging the limitations and potential further study areas.
    
    \subsection{Research Questions}
    \label{research_questions}
    In this sub-section, we aim to address the background, objective, and outcome of conducting the survey.
        \begin{itemize}
            \item \textbf{Q1:} How can control flow graphs (CFG) be used to identify malware derivatives?
            \item \textbf{Q2:} What are the existing machine learning (ML) approaches to analyze malware using CFG?
            \item \textbf{Q3:} What are the drawbacks of existing ML approaches in processing CFG to classify malware?
        \end{itemize}

    \subsection{Inclusion and Criteria}
    In this sub-section, we aim to define the survey's scope by defining the development areas and the filtering process we followed. 
    
    We included a paper if: 
        \begin{itemize}
            \item It contained information relevant to a research question. 
            \item It was written in English.
        \end{itemize}
        We excluded a paper if: 
        \begin{itemize}
            \item It did not address malware analysis by CFG using any ML approaches.
            \item It used any approach other than CFG (e.g., network behavior analysis) with ML to analyze malware.
            \item It used CFG to analyze any behavior (e.g., program characteristics) other than malware.
            \item It was greater than 10 years old.
        \end{itemize}

    \subsection{Data Collection}
    \label{data_collection}
    
    In this sub-section, we list our questions towards collecting information from each piece of literature, described in section \ref{results}.
        \begin{itemize}
            \item What ML framework did the study use to process CFG?
            \item How does the practice affect malware analysis?
            \item What experimental evidence has been provided to support its developmental claims?
        \end{itemize}

    \subsection{Data Analysis}
    This sub-section lists the questions we asked about the data collected by pre-stated [\ref{data_collection}] questions to answer our research objective in section \ref{discussion}:
        \begin{itemize}
            \item How was the ML framework used and what impact does it have on malware analysis using CFG?
            \item How  was  the  research conducted and  analyzed?  Was  it conducted and analyzed reliably and validly?
            \item How does the study relate to other developmental studies? Is it consistent or contradictory?
            \item What claims did the study make on the development?
        \end{itemize}

\section{Results}
\label{results}

    This section summarises our findings and how the study shapes the current malware analysis landscape through CFG using ML. We present different types of malware analysis approaches over time with a summarized Table \ref{tab:cfg_malware_table} and an appendix Table \ref{tab:cfg_appendix_table}. The literature is categorized based on malware platforms in chronological order.

    \subsection{Android Malwares}
    According to Yahoo Finance news, Android is the most popular platform, covering 71.54\% of all smartphones in 2022 \cite{yahoomoney}. In 2017, Symantec intercepted an average of 24000 mobile phone malware per day \cite{ma2019combination}, demonstrating to the requirements of accurate yet efficient malware classification techniques.
    
    % Atici et al. \cite{atici2016android} - Android malware analysis approach based on control flow graphs and machine learning algorithms - 2016
    Therefore, to secure Android devices from malware, Atici et al. \cite{atici2016android} proposed a malware analysis approach using CFG code block grammar. The approach generates CFG from the Android Dalvik byte code instructions. Then, CFG code blocks are represented as string literals— and using string encoding, input vectors from the string literals are generated for the ML algorithms to classify multi-class malware variants. Due to the straightforward approach, the model classification data dimension is reduced to 30 different code chunks, making it fast yet efficient. According to the experiments with the Android Malware Genome Project dataset \cite{zhou2012dissecting}, the model was able to attain a classification accuracy of 96.26\% in general. On top of that, it was able to detect DroidKungfu malware families with a detection rate of 99.15\%; these families are difficult to detect with traditional approaches.

    % Xu et al. \cite{xu2018cdgdroid} - CDGDroid: Android malware detection based on deep learning using CFG and DFG - 2018
    To consider the Android application run-time behaviors with data traffic, Xu et al. \cite{xu2018cdgdroid} proposed another approach (CDGDroid) considering CFG and  data flow graph (DFG). The approach primarily consists of three phases. In the first phase, the CFG and DFG graphs are  extracted. Then, the graphs are encoded for the model to get trained and learn classification. Lastly, the encoded matrix is fed to the deep learning convolutional neural network (CNN) model to learn and detect unseen malicious or normal applications. For extraction and encoding, CFG and DFG graphs are extracted from smali files using Dalvik executions. Then, both graphs are combined via matrix addition or extension to be encoded further. Finally, the encoded matrix is fed to CNN for learning the malware characteristics. Experiments using Marvin \cite{lindorfer2015marvin}, Drebin \cite{arp2014drebin}, VirusShare \cite{virusshare}, and ContagioDump \cite{contagiodump} datasets were conducted. According to the results, the proposed model achieved an accuracy of 99.8\% over Marvin and 72.8\% over the CognitiDump dataset in the detection of unseen malware derivatives. On top of the traditional experiments, a 10-fold cross-validation test was conducted to justify the effectiveness of malware detection using CFG and DFG using deep learning models.

    % Ma et al. \cite{ma2019combination} - A Combination Method for Android Malware Detection Based on Control Flow Graphs and Machine Learning Algorithms - 2019
    To further improve the approach, Ma et al. \cite{ma2019combination} proposed an ensemble ML models approach considering Android API calls, frequency, and sequences. In the approach, the authors constructed a boolean, frequency, and time-series chronological dataset to develop three ML detection models. The diverse API calls and different usage behaviors based on the different attack types is the primary reason behind considering these diversified API usage datasets. First, CFG is constructed from de-complied Android source code. Then, three API datasets (boolean, frequency, and chronological) are constructed from CFG. After that, three ML models dedicated to each dataset type are built for malware analysis and classification. API usage detection model utilizing the boolean dataset is built using a decision tree algorithm. A deep neural network model is used to learn and analyze the API frequency patterns. Lastly, the API sequence detection model is developed using long short-term memory networks (LSTM). Experiments on 10010 benign samples collected from AndroZoo \cite{allix2016androzoo} and 10683 malicious samples collected from Android Malware Dataset \cite{li2017android, wei2017deep} confirmed its detection accuracy of 98.98\%.

    \subsection{Industrial \& IoT Malwares}
    Sophisticated malware like a metamorphic or polymorphic virus can effectively evade signature-based method-based tools by using advanced obfuscation techniques, including mutation and dynamically executed content (DEC) methods. Using DEC, it can dynamically produce new executable code in the run-time, making it difficult to recognize \cite{nguyen2018auto}. According to AV-Test, the total number of malware applications by the end of March 2023 was estimated to be over 1200 million and has increased over 10 times during the last decade \cite{avtest}. Parallelly, IoT devices are especially prone to malware attacks because they are built on light and optimized system architecture prioritizing efficiency. Due to the global IoT adaptation in home and industrial systems, malware developers pay significant attention to disrupting such a landscape resulting in heavy losses. Therefore, polymorphic malware needs to be addressed efficiently for security and economic reasons.

    % Nguyen et al. \cite{nguyen2018auto} - Auto-detection of sophisticated malware using lazy-binding control flow graph and deep learning - 2018
    To capture the malware dynamically executed contents (DEC) behavior, Nguyen et al.\cite{nguyen2018auto} proposed a CFG analysis approach using deep learning. DEC behavior refers to a code obfuscation technique that allows the malware to generate new code at run-time. For such, DEC behaviors are captured in the CFG using lazy binding. Therefore, all the binding between a memory location and corresponding assembly instructions is mapped into the CFG. To extract such CFG, the authors used BE-PUM \cite{nguyen2013hybrid}, which applies on-the-fly push-down model generation from x86 binaries on dynamic symbolic execution in a breadth-first manner. After that, the CFG adjacency matrix is hashed to make it memory efficient by mapping the memory vector to a fixed string length. Finally, the hashed CFG adjacency matrix is fed to CNN to learn the pattern and identify malware. Experiments on real-world samples were collected from VXHeaven \cite{vxheavens}, Virusshare \cite{virusshare}, and MALICIA \cite{nappa2015malicia} datasets containing 63690 malware and 13752 benign programs \cite{younet}. Evaluations using 10-fold cross-validation showcased an average accuracy of over 92\% on all data samples using Yolo-based CNN.
    
    % Yan et al.\cite{yan2019classifying} - Classifying malware represented as control flow graphs using deep graph convolutional neural network - 2019
    To overcome the existing drawbacks of inefficient and ineffective graph mining techniques that commonly rely on handcrafted features and ensemble methods, Yan et al.\cite{yan2019classifying} proposed a malware classification tool that utilizes the graph mining capabilities of Deep Graph Convolution Network (DGCNN). As the CFG follows a heterogeneous data structure, therefore is in tensor of variable size. Hence, it requires a graph machine learning approach to be considered. First, the CFG is extracted using a commercial reverse engineering software called IDA Pro \cite{idapro}. Then, the CFG of unordered size is converted to a fixed size and order. Finally, the CFG tensor is fed to DGCNN to learn to classify using the Adam optimizer. Experiments on MSKCFG \cite{ronen2018microsoft} and YANCFG \cite{yancfg, virustotal} datasets, each containing more than 10000 samples, were conducted and evaluated with 5-fold cross-validation. According to the evaluation, the model achieved an average F-1 score of 0.97 in the MSKCFG dataset and around 0.8 in the YANCFG dataset. Due to the generic approach, the model can be deployed in the cloud for real-time malware classification by a generic user.

    \begin{figure*}[ht]
      \centering
      \includegraphics[width=\textwidth]{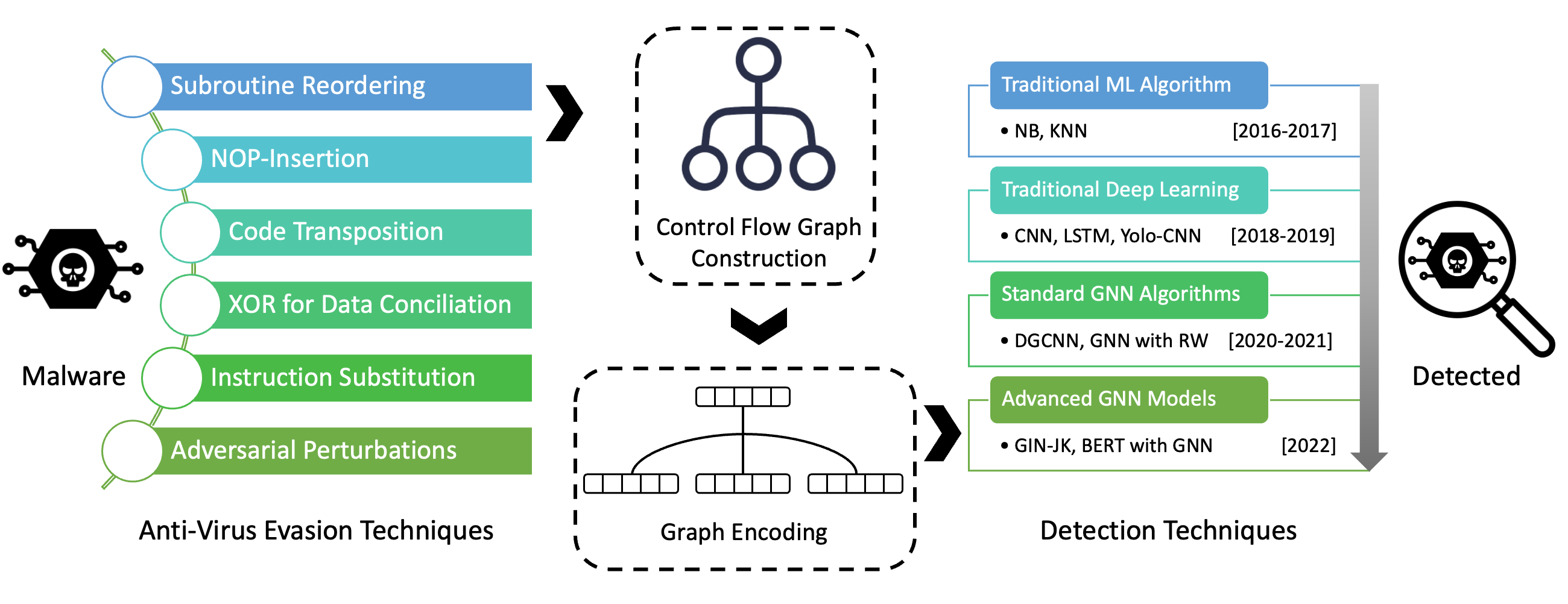}
      \caption{Malware detection life-cycle, from evasion techniques to CFG construction and CFG encoding to detection using ML approaches in chronological order. On the left, obfuscation techniques used by malware are listed, and on the right various ML detection approaches in chronological order are listed.}
      \label{Fig:cfg_diag}
      \vspace{-5.5mm}
    \end{figure*}

    \subsection{Adversarial Malwares}
    The variety and quantity of malware have increased rapidly, complicating classification based on fixed features. Also, the output of an ML model depends on the pattern of the training input samples. Therefore, any unseen pattern can go undetected and evade the anti-virus systems. Apart from numerous code obfuscation techniques, modern malware developers inject adversarial perturbations in the program to make it difficult to detect using standard malware classifiers.

    % Alasmary et al. \cite{alasmary2020soteria} - Soteria: Detecting adversarial examples in control flow graph-based malware classifiers - 2020
     To consider adversarial examples (AEs), Alasmary et al. \cite{alasmary2020soteria} proposed a novel approach (Soteria) that can detect AEs for improved malware classification using deep learning. The model works in two phases: AEs detector and IoT malware classifier. First, the model starts by labeling the extracted CFG nodes by density and level-based labeling. Then, it uses a set of random walk algorithms proportional to the number of nodes in CFG for feature extraction. After that, uses the n-gram module to express and represent the behavior of the software process deeply. Finally, using an auto-encoder, the AEs are detected. The classifier works using an ensemble method of density-based and a level-based CNN classifier. Due to the loosely coupled system architecture and classifier reliance on the AE detector, the classifier does not require extracting features, optimizing the cost. Additionally, the classifier being a separate component allows the user to select a different classifier based on the scenario requirement. An experiment containing randomly selected 13798 malicious samples collected from CyberIOCs \cite{freeioc} was conducted, and feature validation was done using principal component analysis (PCA). Based on the evaluation, the AE detector was able to achieve an accuracy of 97.79\% for detecting AEs, and 99.91\% overall as a multi-class classifier.

     % Hua et al. \cite{hua2020classifying} - Classifying Packed Malware Represented as Control Flow Graphs using Deep Graph Convolutional Neural Network - 2020
     Unlike code obfuscation techniques, packed malware is another approach to bypass malware detection tools. In the approach, malware is unpacked while executing, leading to a different CFG at run-time. Due to the deviated run-time CFG, the malware is able to bypass detection. To address such, Hua et al. \cite{hua2020classifying} proposed an approach to strip the unpacked CFG into local CFG for final classification using DGCNN. In the approach, first, the unpacked CFG is stripped to local CFG by running in a sandbox. The unpack function calls do not relate to any malware local functions and vice-versa. Therefore, from the call adjacency matrix, the local CFG is stripped and generated for classification. Finally, using DGCNN the malicious local CFG is learned for further classification. Experiments covering 6 malware families \cite{virusshare}, each with 100 samples, were conducted using 10-fold cross-validation and were able to demonstrate overall accuracy of 96.4\%. 

     % Wu et al. \cite{wu2021malware} - Malware Classification by Learning Semantic and Structural Features of Control Flow Graphs - 2021
     In order to extend the robustness of the ML models, Wu et al. \cite{wu2021malware} proposed a malware classifier (MCBG) with Graph Isomorphism Network using Jumping Knowledge (GIN-JK). Utilizing the extensive pattern-learning capabilities of modern ML frameworks, the model is capable of learning semantic information about the function nodes as well as the structural information of the entire program CFG. Therefore, adversarial attacks such as code obfuscation or packing can also be considered with the approach. To capture the semantics, it considers the basic program blocks as string literals. Then bidirectional encoder representations from transformers (BERT) is used to pre-train and convert the raw instructions into tokens using masked language model (MLM) and next block prediction (NBP) tasks to generate node-embeddings. Such a pre-trained embedding converts the CFG to attributed CFG (ACFG). Finally, using GIN-JK, the structural information of the program is learned for malware classification. The reason behind selecting GIN-JK with a pooling function for graph representation over traditional GNN is its proven capabilities to allow learning in a simplistic yet efficient manner. Experiments were conducted using Microsoft Malware Classification Challenge (BIG2015) dataset \cite{ronen2018microsoft} with 10868 labeled malware samples of 9 malware families. With a 5-fold cross-validation test set, the model was able to achieve an accuracy of 99.53\%.

     % Herath et al. \cite{herath2022cfgexplainer} - CFGExplainer: Explaining Graph Neural Network-Based Malware Classification from Control Flow Graphs - 2022
     With the diverse ML approaches utilizing GNN to learn the CFG pattern for malware classification, none provide insights into the behavior. To address the explainability while detecting, Herath et al. \cite{herath2022cfgexplainer} proposed a novel DL approach that identifies the most contributing CFG sub-graph alongside malware classification. Such a solution helps security analysts to identify node importance and analyze the behavior in a white-box manner. The model works using two inter-connected feed-forward DNNs. The first component learns to score the node embeddings produced from GNN, and the second component weights the original node embeddings with the produced scores to train a surrogate malware classifier. As both the models are jointly trained using a log-likelihood loss function, contribute to boosting the important node embeddings for malware classification by the second one. Due to the intended objective to learn the CFG node importance, the model is capable of addressing adversarial evasion techniques such as XOR obfuscation, Sematic NOP obfuscation, code manipulation, etc. Experiments with YANCFG dataset \cite{yan2019classifying} over three state-of-the-art models (GNNExplainer \cite{ying2019gnnexplainer}, SubgraphX \cite{yuan2021explainability}, and PGExplainer \cite{luo2020parameterized}) justify the feasibility of the approach.

% #########################################################################

    \begin{table*}[!htbp]
    \caption{Research on CFG Analysis using ML for Malware Analysis}\label{tab:cfg_malware_table}
    \centering
    \setlength{\leftmargini}{0.2cm}
    \bgroup
    \def\arraystretch{1.3}
    % \small
    \begin{tabular}{ |p{3cm}||p{3.5cm}|p{6cm}|p{4cm}|}
    
    \rowcolor{LightCyan}
    
         \hline
         Paper Title & Focus/Objective & Contributions & Limitations\\

         \hline
         \rowcolor{lightgray}
         \multicolumn{4}{|c|}{Android Malwares} \\
         
         \hline
         Android malware analysis approach based on control flow graphs and machine learning algorithms \cite{atici2016android} 
         & Analyze Android malware from code block grammar representation of CFG using ML algorithms.
         & \begin{itemize}
             \item Used CFG code block string representation for multi-class Android malware analysis.
             \item Tested with multiple ML approaches  (Regression Tree, NB, K-NN) to obtain improved accuracy.
         \end{itemize}
         & \begin{itemize}
             \item Might be vulnerable to junk code or code obfuscation techniques and will require improved ML approaches.
         \end{itemize}

         \\
         \hline
         CDGDroid: Android malware detection based on deep learning using CFG and DFG \cite{xu2018cdgdroid} 
         & Malware detection using deep learning over the semantics graph representation of CFG and DFG of Android applications.
         & \begin{itemize}
             \item Used CFG and DFG for malware detection using deep learning techniques.
             \item Used CNN to learn and analyze the malware.
             \item Used 10-fold cross-validation to prove the approach's effectiveness.
         \end{itemize}
         & \begin{itemize}
             \item Better function call graph could have been used.
             \item More experiments are needed targeting malware anti-detection techniques.
         \end{itemize}

         \\
         \hline
         A Combination Method for Android Malware Detection Based on Control Flow Graphs and Machine Learning Algorithms \cite{ma2019combination} 
         & Android malware detection through API calls, API frequency, and API sequence obtained from CFG using an ensemble ML model.
         & \begin{itemize}
             \item Developed malware analysis using API usage, frequency, and sequence detection ensemble model.
             \item Developed chronological API dataset generation method and use LSTM to analyze.
         \end{itemize}
         & \begin{itemize}
             \item The model is not capable to detect malware families.
             \item Using API workflows, the root cause of malware behavior could have been identified.
         \end{itemize}
         \\ 
         
         \hline
         \rowcolor{lightgray}
         \multicolumn{4}{|c|}{Industrial \& IoT Malwares}
         
         \\
         \hline
         Auto-detection of sophisticated malware using lazy-binding control flow graph and deep learning \cite{nguyen2018auto} 
         &  Detect sophisticated malware that employs mutation or dynamically executed content (DEC) behavior using deep learning from CFG graph images generated using lazy binding.
         & \begin{itemize}
             \item Used lazy binding to capture DEC behavior in CFG to detect polymorphic malware.
             \item Used CFG adjacency matrix as an image input for CNN to analyze malware.
         \end{itemize}
         & \begin{itemize}
             \item The CFG generation process is computation heavy.
             \item CNN training requires robust server capabilities with such an approach. 
         \end{itemize}
    
         \\
         \hline
         Classifying malware represented as control flow graphs using deep graph convolutional neural network \cite{yan2019classifying} 
         & Detect malware from CFG using DGCNN to be deployable in a variety of operational environments.
         & \begin{itemize}
             \item Used DGCNN for CFG analysis and classification.
             \item Developed in a generic manner to be deployable in the cloud and can be used by a common user.
         \end{itemize}
         & \begin{itemize}
             \item Requires a high training time.
             \item Requires testing with the latest malware samples for robustness.
         \end{itemize}
         \\

         \hline
         \rowcolor{lightgray}
         \multicolumn{4}{|c|}{Adversarial Malwares}
    
         \\
         \hline
         Soteria: Detecting adversarial examples in control flow graph-based malware classifiers \cite{alasmary2020soteria} 
         & Detect malware while considering adversarial CFG samples using deep learning and random walk algorithms.
         & \begin{itemize}
             \item Developed a model to detect CFG samples targeted for adversarial attacks without training.
             \item Developed multi-class malware classifier using CFG in the IoT domain.
         \end{itemize}
         & \begin{itemize}
             \item AE detector can be bypassed by CFG alteration.
             \item AE detector is susceptible to code obfuscations that result in incomplete CFG.
         \end{itemize}
    
         \\
         \hline
         Classifying Packed Malware Represented as Control Flow Graphs using Deep Graph Convolutional Neural Network \cite{hua2020classifying} 
         & Develop malware classifier using DGCNN while considering the unpacked and local CFG of applications.
         & \begin{itemize}
             \item Developed algorithm to strip from packed CFG to obtain unpacked local CFG.
             \item Used DGCNN to learn and classify malware from unpacked block CFGs. 
         \end{itemize}
         & \begin{itemize}
             \item Different adversarial CFG characteristics are required to be tested with the approach for robust applicability.
         \end{itemize}
    
         \\
         \hline
         Malware Classification by Learning Semantic and Structural Features of Control Flow Graphs \cite{wu2021malware} 
         & Developed a malware classification model by learning the semantic and structural knowledge from CFG using GIN-JK.   
         & \begin{itemize}
             \item Used BERT, MLM, and NBP for pretraining and generating ACFGs with semantic node embedding.
             \item Used GIN-JK to learn CFG and classify malware.
             \item The model is able to handle code obfuscated and packed malware.
         \end{itemize}
         & \begin{itemize}
             \item For packed malware, the model cannot learn the local CFG.
             \item More detailed processing is required on system calls for instruction normalization. 
         \end{itemize}

         \\
         \hline
         CFGExplainer: Explaining Graph Neural Network-Based Malware Classification from Control Flow Graphs \cite{herath2022cfgexplainer} 
         & Developed a deep learning based model to identify CFG sub-graphs that contribute most towards malware classification and justify node importance.
         & \begin{itemize}
             \item Developed DL-based classifier to identify the most contributing CFG sub-graphs for malware classification by GNN models.
             \item Tested model accuracy with three state-of-the-art approaches.
         \end{itemize}
         & \begin{itemize}
             \item Experiment covering more datasets and additional matrices such as sparsity and fidelity are required to evaluate the robustness. 
         \end{itemize}
         \\
         
         \hline
         \rowcolor{lightgray}
         \multicolumn{4}{|c|}{Windows Malwares}
    
         \\
         \hline
         Leveraging Spectral Representations of Control Flow Graphs for Efficient Analysis of Windows Malware \cite{sun2022leveraging} 
         & Developed Windows malware classification using the wave features and heat representations of CFG using NetLSD and spectral graph theory.
         & \begin{itemize}
             \item Used CFG wave features and heat representations for malware classification.
             \item Tested with eight different ML models for evaluation.
         \end{itemize}
         & \begin{itemize}
             \item Further investigation in the spectral representation for malware variants and how perturbations impact these are required.
         \end{itemize}
         
         \\
         \hline
     
    \end{tabular}}
    \end{table*}

     \begin{table*}[!htbp]
        \caption{Research Appendix}\label{tab:cfg_appendix_table}
        \centering
        \setlength{\leftmargini}{0.2cm}
        \bgroup
        \def\arraystretch{1.3}
        % \small
        \begin{tabular}{ |p{2.3cm}||p{0.7cm}|p{2.5cm}|p{3cm}|p{3cm}|p{3cm}|}
        
        \rowcolor{LightCyan}
        
         \hline
         Paper & Year & CFG-Extraction Tools & ML Models & Datasets & Performance\\

         \hline
         \rowcolor{lightgray}
         \multicolumn{6}{|c|}{Android Malwares} \\

              \hline
              Atici et al. \cite{atici2016android} &
              2016 &
              Androguard \cite{androguard} &
              CART (Classification And Regression Tree) \cite{rutkowski2014cart} &
              Android Malware Genome Project \cite{zhou2012dissecting} & 
              96.26\% - Accuracy\\
    
              \hline
              Xu et al. \cite{xu2018cdgdroid} &
              2018 &
              Apktool \cite{apktool} &
              CNN & 
              Marvin \cite{lindorfer2015marvin}, Drebin \cite{arp2014drebin}, VirusShare \cite{virusshare}, ContagioDump \cite{contagiodump}, Mi App Store \cite{miappstore}, VirusTotal \cite{virustotal} &
              Marvin: 
              \newline 
              99.822\% - Accuracy
              \newline
              99.191\% - F1 
              \newline
              ContagioDump:
              \newline
              72.87\% - Accuracy
              \newline
              84.301\% - F1
              \\
    
              \hline
              Ma et al. \cite{ma2019combination} &
              2019 &
              Modified FlowDroid \cite{flowdroid} &
              Decision Tree, DNN \& LSTM - Ensemble &
              AndroZoo \cite{allix2016androzoo}, Android Malware Dataset \cite{li2017android, wei2017deep} & 
              98.98\% - F1\\

         \hline
         \rowcolor{lightgray}
         \multicolumn{6}{|c|}{Industrial \& IoT Malwares}\\

             \hline
              Nguyen et al. \cite{nguyen2018auto} &
              2018 &
              BE-PUM \cite{nguyen2013hybrid} &
              Simple CNN \& Yolo-based CNN &
              VXHeaven \cite{vxheavens}, VirusShare \cite{virusshare}, MALICIA \cite{nappa2015malicia} & 
              92.53 - F1 (Yolo CNN)\\
    
              \hline
              Yan et al.\cite{yan2019classifying} &
              2019 &
              IDA Pro \cite{idapro} &
              DGCNN &
              Microsoft Malware Classification Challange \cite{ronen2018microsoft}, YANCFG \cite{yancfg, virustotal} & 
              97\% - F1\\

         \hline
         \rowcolor{lightgray}
         \multicolumn{6}{|c|}{Adversarial Malwares}\\

             \hline
              Alasmary et al. \cite{alasmary2020soteria} &
              2020 &
              Radare2 \cite{radare2} &
              CNN &
              CyberIOCs \cite{freeioc} & 
              99.91\% - Accuracy\\
    
              \hline
              Hua et al. \cite{hua2020classifying} &
              2020 &
              Sandbox &
              DGCNN &
              VirusShare \cite{virusshare} & 
              96.4\% - Accuracy \\
    
              \hline
              Wu et al. \cite{wu2021malware} &
              2021 &
              IDA Pro \cite{idapro}, Radare2 \cite{radare2} &
              BERT \cite{devlin2018bert}, GIN-JK \cite{xu2018representation} &
              Microsoft Malware Classification Challange \cite{ronen2018microsoft} & 
              99.53\% - Accuracy \\
    
              \hline
              Herath et al. \cite{herath2022cfgexplainer} &
              2022 &
              IDA Pro \cite{idapro}, Ghidra \cite{ghidra} &
              GNN &
              YANCFG \cite{yan2019classifying} & 
              0.8018 - AUC\\

         \hline
         \rowcolor{lightgray}
         \multicolumn{6}{|c|}{Windows Malwares}\\

             \hline
              Sun et al. \cite{sun2022leveraging} &
              2022 &
              Radare2 - r2pipe \cite{radare2} &
              SVM, DT, LR, RF, KNN, ANN, AdaBoost, XGB &
              Windows malware samples \cite{aghakhani2020malware} & 
              95.9\% - Accuracy \\
         
         \hline
        \end{tabular}}
    \end{table*}

% ############################################################

     \subsection{Windows Malwares}
     Windows is the most globally used operating system, making it an important playground for malware developers to target general users. For example, recent Ransomware attacks incurred a heavy cost to general users. On top of that, anti-malware tools are vulnerable to even general code obfuscation techniques, and 90\% of the signature-based approaches don't conduct other static analysis \cite{atici2016android}.

     % Sun et al. \cite{sun2022leveraging} - Leveraging Spectral Representations of Control Flow Graphs for Efficient Analysis of Windows Malware - 2022
     Due to the high customer base of Windows OS, Sun et al. \cite{sun2022leveraging} proposed a novel approach to analyze the CFG wave features and heat representations with ML models for malware classification. In the poster work, the authors tested eight ML models (SVM, DT, LR, RF, KNN, ANN, AdaBoost, and XGB) over CFG wave and heat notations to identify the best approach. The reason behind considering wave feature and heat representation is the size efficient and permutation invariant characteristics of such. First, using r2pipe API, CFG was extracted and constructed 250 to 1000-dimensional heat and wave spectral graphs using NetLSD \cite{tsitsulin2018netlsd}. NetLSD offers to generate compact graph signatures using Laplacian heat or wave kernel inheriting Laplacian spectrum's formal features. Also, PCA was used for dimensionality reduction of 250 to 1000 features, to make the model efficient. To conduct the experiment, the authors used 37537 Windows malware samples \cite{aghakhani2020malware} with a 70\% to 30\% train test split ratio. According to the experiment, the wave features were the most accurate to classify malware using RF, DT, and XGB with a maximum accuracy of 95.9\% compared to heat representations.

\section{Discussion}
\label{discussion}

    In this section, we discuss the answers to our research questions stated in section [\ref{research_questions}] based on the findings discussed in the result section [\ref{results}]. Each subsection is addressing each of our research questions with limitations. We also illustrate the evolving research landscape through Fig \ref{Fig:cfg_diag}.

    \subsection{Evolution of CFG Analysis}
    To protect legitimate users from malicious threats, many approaches have been proposed over the last decade. For real-time protection, control flow graph analysis is one of the prominent. A control flow graph reflects the intended program behavior as a graph, allowing malicious behavioral patterns to be captured. \\
    In the initial studies, CFG code blocks were encoded as string literal for program behavior pattern identification using NLP techniques as demonstrated by Atici et al.\cite{atici2016android}. Using such an approach, the underlying malicious behavioral logic was neglected to identify patterns. Later, to focus more on the available graph data, DFG was also considered for better pattern identification alongside CFG, as demonstrated by Xu et al. \cite{xu2018cdgdroid}. With increased data to process, the proportional increase in computation time also became a limiting factor. Furthermore, numerous ensemble models utilizing different CFGs were considered to attain increased accuracy \cite{ma2019combination}. However, the program logic being the primary factor determining malicious behavior, such approaches were inefficient to detect evasion techniques like NOP insertion, code transposition, etc. Therefore, in order to consider the instruction node semantics, CFG encoding was necessary for detailed analysis as demonstrated by Yan et al. \cite{yan2019classifying}. Apart from typical code obfuscation techniques, malware developers adopted different adversarial methods like Packing. Therefore, on top of the spectral features, the program's structural flow demanded to be analyzed as shown by Wu et al. \cite{wu2021malware}. Apart from traditional techniques, to keep up with the regularly evolving cyber threat landscape, researchers also tested a few unconventional approaches like encoding CFG as an image \cite{nguyen2018auto}, analyzing CFG heat representation and wave features \cite{sun2022leveraging}, etc., for classification. 

    \subsection{Evolving ML Approaches}
    Traditionally, CFG analysis was conducted by security analysts for malware classification in security centers. With the evolution of ML, now models are getting trained to identify such patterns in no time with impressive accuracy while reducing analysis costs.\\
    In the initial CFG analysis approaches, typical NLP techniques or basic ML algorithms like KNN, NB, Regression Tree, etc., were used to allow the model to be accurate yet efficient, as demonstrated in \cite{atici2016android}. Due to limited algorithmic capabilities, the models were unable to address the underlying CFG logic. Hence, to capture the underlying instruction pattern, researchers used various CNN models as demonstrated by Yan et al. \cite{yan2019classifying}. However, the incompatibility of CNN models with large heterogeneous CFG structures leads researchers to adopt GNN models for pattern analysis. Being a relatively new ML approach, numerous research approaches with various GNN models like GCN, GIN-JK, etc., are being carried out to target improvement areas; studies by Wu et al. \cite{wu2021malware}, Herath et al. \cite{herath2022cfgexplainer} are a few.

    \subsection{Limitations and Evolution}
    With increasing computational power and memory, program dimensions are also increasing in proportion. Therefore, identifying detailed program behavioral patterns in a limited time with existing deep-learning approaches is becoming difficult. \\
    The three key areas that limit the process are:
    \begin{itemize}
        \item CFG Extraction
        \item Robust CFG Encoding
        \item Pattern Identification \& Explainability
    \end{itemize}
    
    For a device with average computation capabilities like IoT and Android, CFG extraction becomes a heavy load. In many cases, malware can execute malicious activities before being detected by process-heavy models. Therefore, for IoT or Android domains, an efficient CFG extractor development will drastically reduce identification time. 
    
    Secondly, CFG encoding is the main area that enables a model to learn program patterns. The absence of a robust encoding mechanism is the leading cause behind the evasion of unseen malware family derivatives. Therefore, automated feature analysis is required to make a robust encoding standard, allowing the model to comprehend unseen patterns from large datasets. 
    
    Finally, pattern identification is the main task of an anti-malware model. As ML architecture is the main component behind pattern learning, it cannot utilize the encoded information from CFG without a well-designed ML model. As of today, GNN is the most suitable ML model, capable of learning the CFG pattern better than other ML models. But it consumes higher computation time for real-time applicability standards. Therefore, further research with GNN is required to make the model robust yet equally efficient \cite{bilot2023survey}. On top of that, the malware detection models can be incorporated with the cybersecurity knowledge graph (CKG), which is also an active area of research, for in-depth pattern learning from a diverse, authentic data source \cite{mitra2021combating}. Such an outcome would allow the models to learn from the global data repository and is expected to reduce the computation time significantly. \\
    To address all these factors, the root cause of any model behavior must be identified and explained for a production-ready system, which is also an active area of research. Specifically, ML models are prone to adversarial attacks. There have been numerous studies \cite{zhang2022semantics} on conducting malicious attacks on ML models for evasion. Hence, developing a secured ML model is equally important as an efficient one. The study by Herath et al. \cite{herath2022cfgexplainer} is notable example among a few. More research is required in the explainability domain to enable analysis in a directed rather than a trial-and-error manner while considering ML models as the black box.

\section*{Conclusion}

    We have surveyed some of the recent techniques in malware detection through CFGs using ML that have shown significant potential in addressing the limitations of traditional signature-based malware detection methods, highlighting the different aspects of feature extraction, representation, and classification. We have discussed the different types of CFG features that have been used, as well as the different ML algorithms that have been applied to CFG-based malware detection. We have also discussed several challenges and limitations of these methods, such as scalability, robustness, and interpretability, and proposed possible solutions and directions for future research. Specifically, we identified three critical open areas that need further extensive research: the following.
    \begin{itemize}
        \item Effective and efficient CFG extraction.
        \item Robust and accurate ML algorithm to handle large data.
        \item Explainability of ML model behaviors for directed research and secure deployment.
    \end{itemize}
   Overall, we believe CFG-based malware detection using ML is a promising new approach that can provide a high level of accuracy and generality to overcome the limitations of signature-based detection approaches in securing computer systems and networks against the evolving threat of malware.

\section*{Future Work}

    We hope this survey can serve as a helpful reference for researchers and practitioners interested in this field and inspire further developments and innovations in this area. As a future step, we plan to conduct experimental research to address the areas of improvement and persisting loopholes discussed above.

\section*{Acknowledgment}
The author would like to thank Amy Barton for constructive criticism of the manuscript.

% \nocite{*}
\bibliographystyle{unsrt}
\bibliography{conference_101719}

\begin{thebibliography}{10}

\bibitem{kitchenham2004procedures}
Barbara Kitchenham.
\newblock Procedures for performing systematic reviews.
\newblock {\em Keele, UK, Keele University}, 33(2004):1--26, 2004.

\bibitem{yahoomoney}
Available [Online]~Yahoo News.
\newblock accessed: 2023-04-20.
\newblock
  {https://money.yahoo.com/android-dominates-globally-but-apples-gaining-ground-102346830.html}.

\bibitem{ma2019combination}
Zhuo Ma, Haoran Ge, Yang Liu, Meng Zhao, and Jianfeng Ma.
\newblock A combination method for android malware detection based on control
  flow graphs and machine learning algorithms.
\newblock {\em IEEE access}, 7:21235--21245, 2019.

\bibitem{atici2016android}
Mehmet~Ali Atici, Seref Sagiroglu, and Ibrahim~Alper Dogru.
\newblock Android malware analysis approach based on control flow graphs and
  machine learning algorithms.
\newblock In {\em 2016 4th International Symposium on Digital Forensic and
  Security (ISDFS)}, pages 26--31. IEEE, 2016.

\bibitem{zhou2012dissecting}
Yajin Zhou and Xuxian Jiang.
\newblock Dissecting android malware: Characterization and evolution.
\newblock In {\em 2012 IEEE symposium on security and privacy}, pages 95--109.
  IEEE, 2012.

\bibitem{xu2018cdgdroid}
Zhiwu Xu, Kerong Ren, Shengchao Qin, and Florin Craciun.
\newblock Cdgdroid: Android malware detection based on deep learning using cfg
  and dfg.
\newblock In {\em Formal Methods and Software Engineering: 20th International
  Conference on Formal Engineering Methods, ICFEM 2018, Gold Coast, QLD,
  Australia, November 12-16, 2018, Proceedings 20}, pages 177--193. Springer,
  2018.

\bibitem{lindorfer2015marvin}
Martina Lindorfer, Matthias Neugschwandtner, and Christian Platzer.
\newblock Marvin: Efficient and comprehensive mobile app classification through
  static and dynamic analysis.
\newblock In {\em 2015 IEEE 39th annual computer software and applications
  conference}, volume~2, pages 422--433. IEEE, 2015.

\bibitem{arp2014drebin}
Daniel Arp, Michael Spreitzenbarth, Malte Hubner, Hugo Gascon, Konrad Rieck,
  and CERT Siemens.
\newblock Drebin: Effective and explainable detection of android malware in
  your pocket.
\newblock In {\em Ndss}, volume~14, pages 23--26, 2014.

\bibitem{virusshare}
Available~[Online] VirusShare.
\newblock accessed: 2023-04-20.
\newblock {https://virusshare.com/}.

\bibitem{contagiodump}
Available~[Online] Contagio.
\newblock accessed: 2023-04-20.
\newblock {https://contagiodump.blogspot.com/}.

\bibitem{allix2016androzoo}
Kevin Allix, Tegawend{\'e}~F Bissyand{\'e}, Jacques Klein, and Yves Le~Traon.
\newblock Androzoo: Collecting millions of android apps for the research
  community.
\newblock In {\em Proceedings of the 13th international conference on mining
  software repositories}, pages 468--471, 2016.

\bibitem{li2017android}
Yuping Li, Jiyong Jang, Xin Hu, and Xinming Ou.
\newblock Android malware clustering through malicious payload mining.
\newblock In {\em Research in Attacks, Intrusions, and Defenses: 20th
  International Symposium, RAID 2017, Atlanta, GA, USA, September 18--20, 2017,
  Proceedings}, pages 192--214. Springer, 2017.

\bibitem{wei2017deep}
Fengguo Wei, Yuping Li, Sankardas Roy, Xinming Ou, and Wu~Zhou.
\newblock Deep ground truth analysis of current android malware.
\newblock In {\em Detection of Intrusions and Malware, and Vulnerability
  Assessment: 14th International Conference, DIMVA 2017, Bonn, Germany, July
  6-7, 2017, Proceedings 14}, pages 252--276. Springer, 2017.

\bibitem{nguyen2018auto}
Minh~Hai Nguyen, Dung Le~Nguyen, Xuan~Mao Nguyen, and Tho~Thanh Quan.
\newblock Auto-detection of sophisticated malware using lazy-binding control
  flow graph and deep learning.
\newblock {\em Computers \& Security}, 76:128--155, 2018.

\bibitem{avtest}
Available~[Online] AV-test.
\newblock accessed: 2023-04-04.
\newblock {https://www.av-test.org/en/statistics/malware/}.

\bibitem{nguyen2013hybrid}
Minh~Hai Nguyen, Thien~Binh Nguyen, Thanh~Tho Quan, and Mizuhito Ogawa.
\newblock A hybrid approach for control flow graph construction from binary
  code.
\newblock In {\em 2013 20th Asia-Pacific Software Engineering Conference
  (APSEC)}, volume~2, pages 159--164. IEEE, 2013.

\bibitem{vxheavens}
Available~[Online] VXHeavens.
\newblock accessed: 2023-04-20.
\newblock {https://archive.org/download/vxheavens-2010-05-18}.

\bibitem{nappa2015malicia}
Antonio Nappa, M~Zubair Rafique, and Juan Caballero.
\newblock The malicia dataset: identification and analysis of drive-by download
  operations.
\newblock {\em International Journal of Information Security}, 14:15--33, 2015.

\bibitem{younet}
Available~[Online] YouNet.
\newblock accessed: 2023-04-20.
\newblock {https://www.younetgroup.com/}.

\bibitem{yan2019classifying}
Jiaqi Yan, Guanhua Yan, and Dong Jin.
\newblock Classifying malware represented as control flow graphs using deep
  graph convolutional neural network.
\newblock In {\em 2019 49th annual IEEE/IFIP international conference on
  dependable systems and networks (DSN)}, pages 52--63. IEEE, 2019.

\bibitem{idapro}
hex rays.
\newblock Ida pro [online].
\newblock {https://hex-rays.com/ida-pro/}.

\bibitem{ronen2018microsoft}
Royi Ronen, Marian Radu, Corina Feuerstein, Elad Yom-Tov, and Mansour Ahmadi.
\newblock Microsoft malware classification challenge.
\newblock {\em arXiv preprint arXiv:1802.10135}, 2018.

\bibitem{yancfg}
Available~[Online] offensivecomputing.
\newblock accessed: 2023-04-20.
\newblock {http://www.offensivecomputing.net}.

\bibitem{virustotal}
Available~[Online] VirusTotal.
\newblock accessed: 2023-04-20.
\newblock {https://https://www.virustotal.com/}.

\bibitem{alasmary2020soteria}
Hisham Alasmary, Ahmed Abusnaina, Rhongho Jang, Mohammed Abuhamad, Afsah Anwar,
  DaeHun Nyang, and David Mohaisen.
\newblock Soteria: Detecting adversarial examples in control flow graph-based
  malware classifiers.
\newblock In {\em 2020 IEEE 40th International Conference on Distributed
  Computing Systems (ICDCS)}, pages 888--898. IEEE, 2020.

\bibitem{freeioc}
Available~[Online] freeioc.
\newblock accessed: 2019.
\newblock {https://freeiocs.cyberiocs.pro/}.

\bibitem{hua2020classifying}
Yakang Hua, Yuanzheng Du, and Dongzhi He.
\newblock Classifying packed malware represented as control flow graphs using
  deep graph convolutional neural network.
\newblock In {\em 2020 International Conference on Computer Engineering and
  Application (ICCEA)}, pages 254--258. IEEE, 2020.

\bibitem{wu2021malware}
Bolun Wu, Yuanhang Xu, and Futai Zou.
\newblock Malware classification by learning semantic and structural features
  of control flow graphs.
\newblock In {\em 2021 IEEE 20th International Conference on Trust, Security
  and Privacy in Computing and Communications (TrustCom)}, pages 540--547.
  IEEE, 2021.

\bibitem{herath2022cfgexplainer}
Jerome~Dinal Herath, Priti~Prabhakar Wakodikar, Ping Yang, and Guanhua Yan.
\newblock Cfgexplainer: Explaining graph neural network-based malware
  classification from control flow graphs.
\newblock In {\em 2022 52nd Annual IEEE/IFIP International Conference on
  Dependable Systems and Networks (DSN)}, pages 172--184. IEEE, 2022.

\bibitem{ying2019gnnexplainer}
Zhitao Ying, Dylan Bourgeois, Jiaxuan You, Marinka Zitnik, and Jure Leskovec.
\newblock Gnnexplainer: Generating explanations for graph neural networks.
\newblock {\em Advances in neural information processing systems}, 32, 2019.

\bibitem{yuan2021explainability}
Hao Yuan, Haiyang Yu, Jie Wang, Kang Li, and Shuiwang Ji.
\newblock On explainability of graph neural networks via subgraph explorations.
\newblock In {\em International Conference on Machine Learning}, pages
  12241--12252. PMLR, 2021.

\bibitem{luo2020parameterized}
Dongsheng Luo, Wei Cheng, Dongkuan Xu, Wenchao Yu, Bo~Zong, Haifeng Chen, and
  Xiang Zhang.
\newblock Parameterized explainer for graph neural network.
\newblock {\em Advances in neural information processing systems},
  33:19620--19631, 2020.

\bibitem{sun2022leveraging}
Qirui Sun, Eldor Abdukhamidov, Tamer Abuhmed, and Mohammed Abuhamad.
\newblock Leveraging spectral representations of control flow graphs for
  efficient analysis of windows malware.
\newblock In {\em Proceedings of the 2022 ACM on Asia Conference on Computer
  and Communications Security}, pages 1240--1242, 2022.

\bibitem{androguard}
Anthony Desnos and Open~Source Team.
\newblock Androguard [online].
\newblock {https://github.com/androguard/androguard}.

\bibitem{rutkowski2014cart}
Leszek Rutkowski, Maciej Jaworski, Lena Pietruczuk, and Piotr Duda.
\newblock The cart decision tree for mining data streams.
\newblock {\em Information Sciences}, 266:1--15, 2014.

\bibitem{apktool}
Wiśniewski Ryszard and Tumbleson Connor.
\newblock Apktool: A tool for reverse engineering android apk files [online].
\newblock {https://ibotpeaches.github.io/Apktool/}.

\bibitem{miappstore}
Mi.
\newblock Mi app store [online].
\newblock {https://www.dev.mi.com/en}.

\bibitem{flowdroid}
Steven Arzt and Open source team.
\newblock Flowdroid data flow analysis tool [online].
\newblock {https://github.com/secure-software-engineering/FlowDroid}.

\bibitem{radare2}
Radare2.
\newblock Radare2 [online].
\newblock {https://rada.re/n/radare2.html}.

\bibitem{devlin2018bert}
Jacob Devlin, Ming-Wei Chang, Kenton Lee, and Kristina Toutanova.
\newblock Bert: Pre-training of deep bidirectional transformers for language
  understanding.
\newblock {\em arXiv preprint arXiv:1810.04805}, 2018.

\bibitem{xu2018representation}
Keyulu Xu, Chengtao Li, Yonglong Tian, Tomohiro Sonobe, Ken-ichi Kawarabayashi,
  and Stefanie Jegelka.
\newblock Representation learning on graphs with jumping knowledge networks.
\newblock In {\em International conference on machine learning}, pages
  5453--5462. PMLR, 2018.

\bibitem{ghidra}
Ghidra.
\newblock Ghidra [online].
\newblock {https://ghidra-sre.org/}.

\bibitem{aghakhani2020malware}
Hojjat Aghakhani, Fabio Gritti, Francesco Mecca, Martina Lindorfer, Stefano
  Ortolani, Davide Balzarotti, Giovanni Vigna, and Christopher Kruegel.
\newblock When malware is packin'heat; limits of machine learning classifiers
  based on static analysis features.
\newblock In {\em Network and Distributed Systems Security (NDSS) Symposium
  2020}, 2020.

\bibitem{tsitsulin2018netlsd}
Anton Tsitsulin, Davide Mottin, Panagiotis Karras, Alexander Bronstein, and
  Emmanuel M{\"u}ller.
\newblock Netlsd: hearing the shape of a graph.
\newblock In {\em Proceedings of the 24th ACM SIGKDD International Conference
  on Knowledge Discovery \& Data Mining}, pages 2347--2356, 2018.

\bibitem{bilot2023survey}
Tristan Bilot, Nour~El Madhoun, Khaldoun~Al Agha, and Anis Zouaoui.
\newblock A survey on malware detection with graph representation learning.
\newblock {\em arXiv preprint arXiv:2303.16004}, 2023.

\bibitem{mitra2021combating}
Shaswata Mitra, Aritran Piplai, Sudip Mittal, and Anupam Joshi.
\newblock Combating fake cyber threat intelligence using provenance in
  cybersecurity knowledge graphs.
\newblock In {\em 2021 IEEE International Conference on Big Data (Big Data)},
  pages 3316--3323. IEEE, 2021.

\bibitem{zhang2022semantics}
Lan Zhang, Peng Liu, Yoon-Ho Choi, and Ping Chen.
\newblock Semantics-preserving reinforcement learning attack against graph
  neural networks for malware detection.
\newblock {\em IEEE Transactions on Dependable and Secure Computing},
  20(2):1390--1402, 2022.

\end{thebibliography}

\end{document}